# Tropopause occurs due to the general circulation of the atmosphere

**Kochin A.V.**

Central Aerological Observatory

3 Pervomayskaya str., Dolgoprudny, Moscow region, 141707 Russia

Email: amarl@mail.ru

**Abstract**

The general circulation of the atmosphere determines the long-term variability of weather processes. This circulation is driven by the temperature differences between the poles and the equator, causing air to move along the Earth's surface. However, this requires enhanced pressure at the poles, which is not observed. To sustain the circulation, an additional non-hydrostatic pressure gradient is required. Here I propose the emergence of an additional non-hydrostatic pressure gradient resulting from the centrifugal force generated by the Earth's rotation. This centrifugal force creates a non-hydrostatic vertical pressure gradient, which is essential for the closed circulation of unequally heated air in the meridional direction. The circulation is composed of three distinct streams flowing in opposite directions, with the polar and tropical tropopause acting as boundaries. The temperature in the atmosphere decreases from the surface to the polar tropopause and remains constant above it.

**Plain language summary**

Forecasting the characteristics of the general atmospheric circulation is important for long-term weather forecasting in order to ensure energy security. In 1753, the English scientist Hadley proposed a single-cell structure of the general atmospheric circulation, in which air rises at the equator and descends at the poles. However, temperature decrease should lead to increased surface pressure at the poles to drive the air masses along the Earth's surface from poles to equator. The observed surface pressure remains relatively constant, which means that the air can only move towards the poles due to the hydrostatic pressure gradient. To sustain the circulation, an additional non-hydrostatic pressure gradient is required. I consider the process of generating this additional non-hydrostatic pressure gradient through the centrifugal force resulting from the Earth's rotation. The centrifugal force balances the surface pressure and creates a non-hydrostatic vertical pressure gradient necessary for the closed circulation of unequally heated air in the meridional direction. The circulation consists of three oppositely directed streams separated by the polar and tropical tropopause, with the temperature in the atmosphere falling from the surface to the polar tropopause and constant above it.

## 1. Introduction

The general circulation of the atmosphere (GCA) plays a crucial role in transporting heat and moisture between the equator and the poles, thereby influencing long-term atmospheric processes.

It is important to note that the speed of air flows in the GCA is significantly slower compared to the flows involved in mesoscale processes. Cyclones and anticyclones are responsible for more intense heat and water vapor transfer over the Earth's surface, but their effects are temporary and occur in multiple directions. In contrast, the GCA facilitates a permanent and unidirectional transfer, ensuring a sustained and consistent redistribution of heat and moisture across the globe.

The general circulation of the atmosphere is primarily driven by the temperature contrast between the equator and the poles. In colder air, there is a greater vertical pressure gradient, resulting in a pressure difference between warm and cold regions that increases with altitude (Chamberlain 1978, Eckart 1960, Holton J. 2004, Perevedentsev 2013). In order for circulation to occur, it is essential that the atmosphere contains regions with varying directional gradients. As a result, the pressure near the Earth's surface in the vicinity of the poles should be greater than the pressure at the equator. This configuration prompts air near the surface to flow towards the equator, while above a certain height, the flow reverses with a zero gradient towards the poles.

The rotation of the Earth may itself be the cause of circulation in the absence of a temperature gradient. At the equator, the acceleration of gravity decreases due to centrifugal force, which leads to a decrease in the vertical pressure gradient. There is a pressure difference between the equator and the pole, which increases with altitude. This difference can cause the appearance of circulation at high altitudes in the absence of a temperature gradient in the atmosphere.

The meridional flow from the pole to the equator has a zonal component opposite in sign in the flow from the equator to the pole due to the action of the Coriolis force. Under conditions close to calm, a change in the direction of zonal flows makes it possible to detect the boundary between meridional flows.

## 2. Experimental data on vertical temperature and wind profiles.

Detecting the meridional flow velocity in the GCA using current methods is challenging due to its relatively low magnitude, which is not prominently displayed in the wind field. Furthermore, this velocity is significantly smaller compared to speeds observed in mesoscale processes. Consequently, under conditions of relative atmospheric calm, only changes in wind direction within zonal flows can be detected. Fig. 1 presents data from upper-air sounding, illustrating sharp wind direction changes occurring in both the polar and tropical tropopause regions.

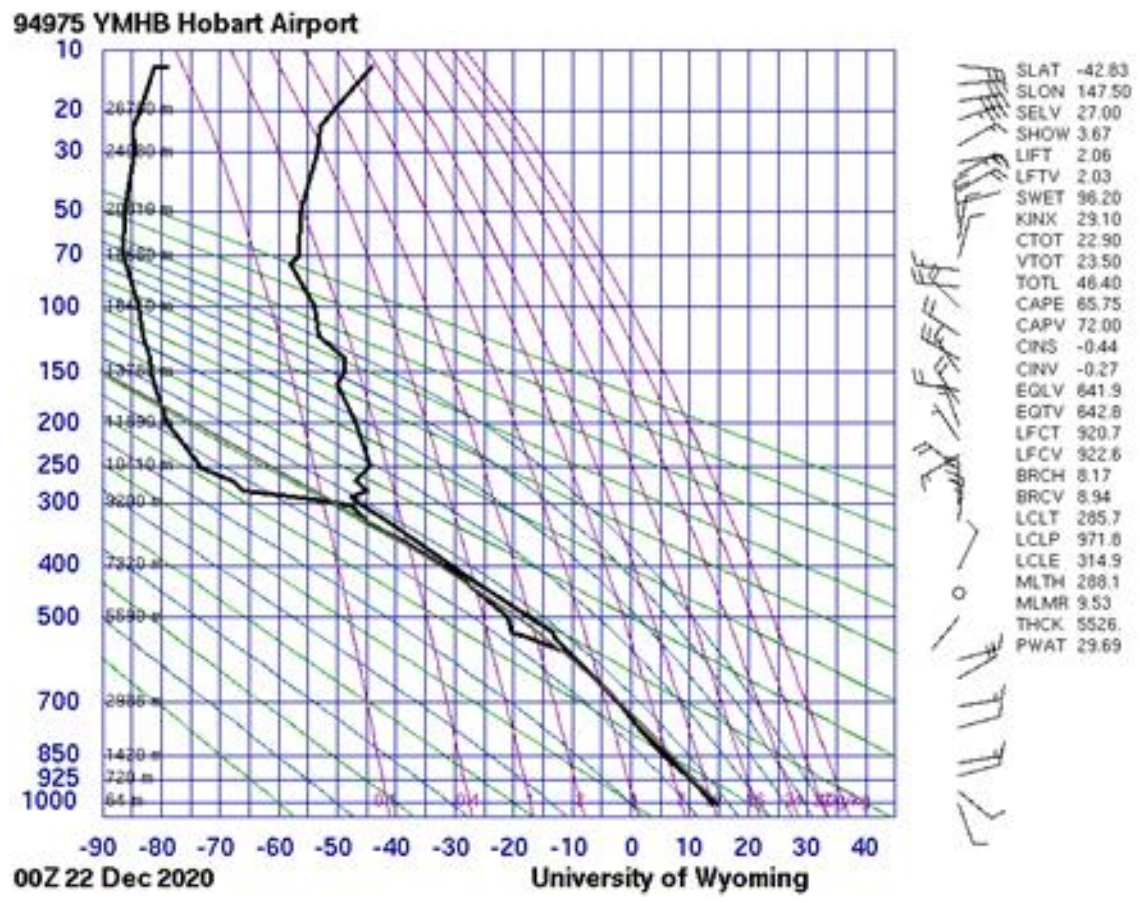

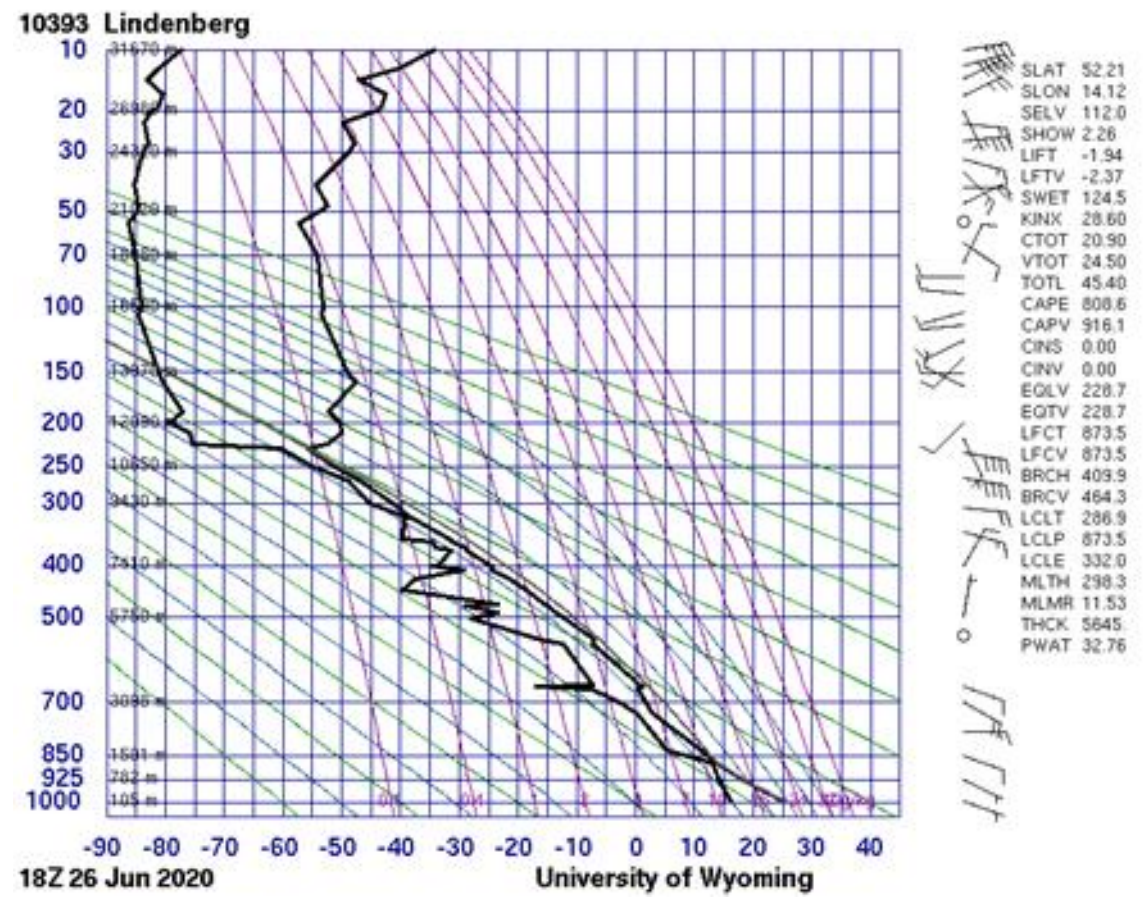

a) Hobart Airport, Australia, Lat: - 43º  b) Lindenberg, Germany, Lat: 52º

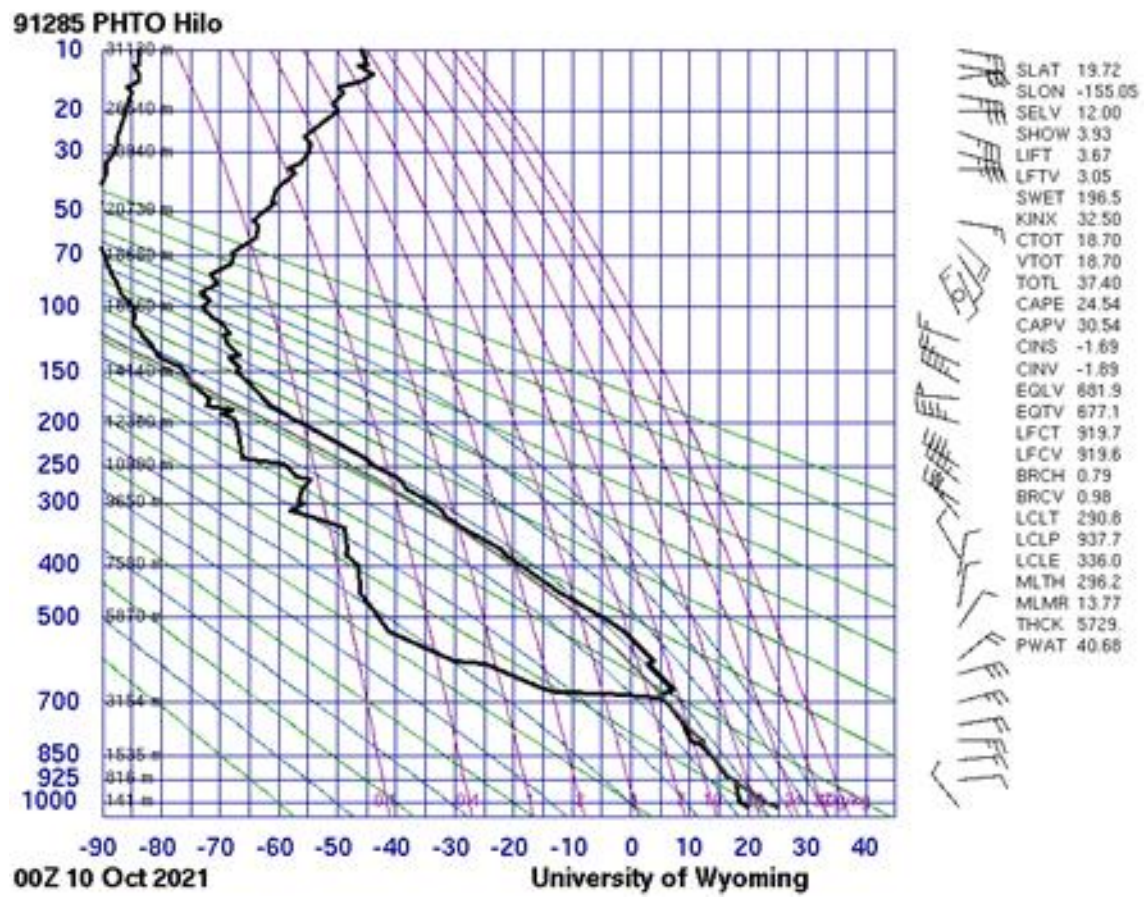

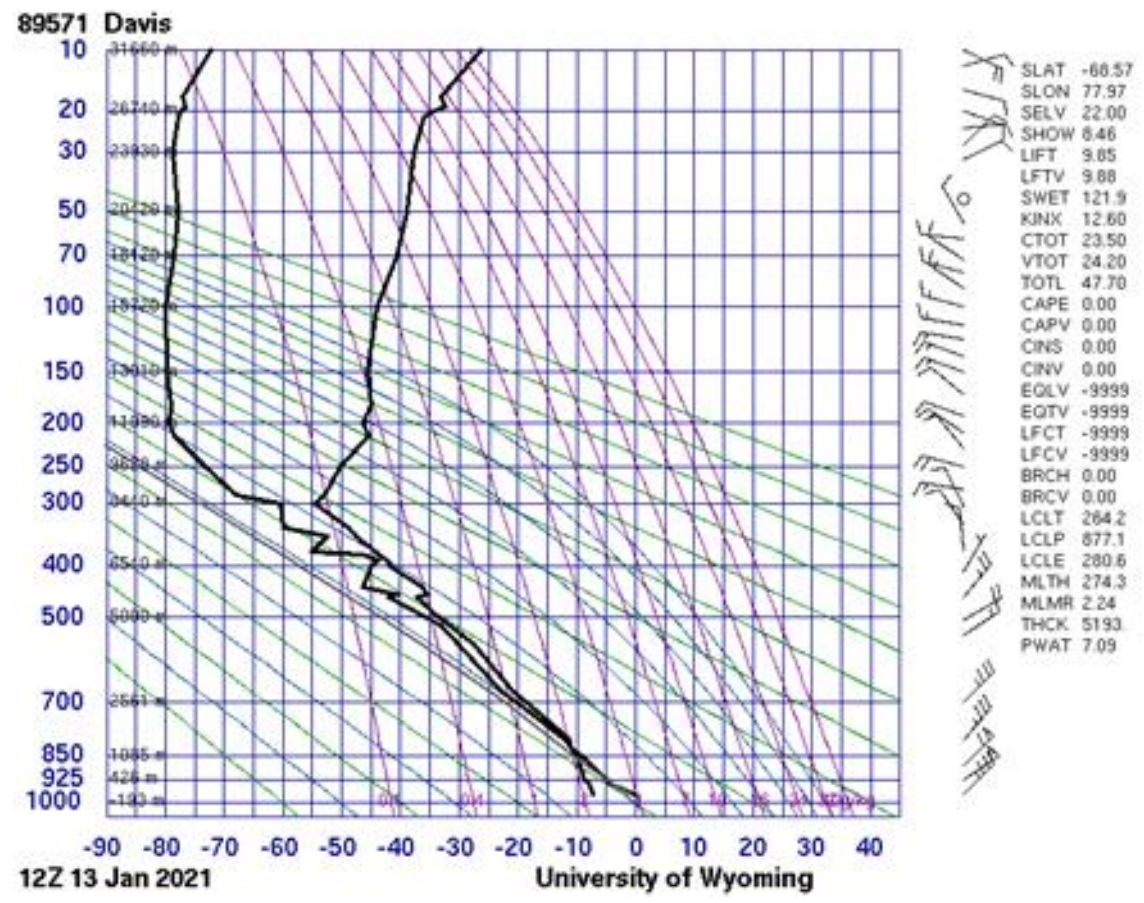

c) Hawaii, Latitude: 20º                                d) Davis, Antarctica, Lat: -69º

*Fig.1. Upper-air sounding results selected by the presence of a sharp change in wind direction near the tropopause. Wind direction is shown by arrows to the right of the charts. The arrow indicates where the wind is blowing. The air temperature is shown as a line closer to the center of the graph. The line on the left is the dew point temperature. The vertical axis shows altitude (m) and pressure (HPa), the horizontal axis shows temperature (Cº). The names of the stations and their latitudes are indicated below the charts (http://weather.uwyo.edu/upperair/sounding.html).*

From the Earth's surface to the tropopause, there is a predominant eastward transport of air (eastward flow). However, between the polar and tropical tropopauses, the flow direction reverses, resulting in a westward transport (western flow). Above the tropical tropopause, the air once again moves eastward (eastward transport). Consequently, the meridional transfer of air near the Earth's surface and above the tropical tropopause is directed towards the equator, while between the tropopauses, it is directed towards the pole. It turns out that there is a second circulation cell around the tropical tropopause, and the area between the tropopause is a hollow sphere that rotates from west to east.

3. **Formation of the general circulation of the atmosphere due to the rotation of the Earth.**

Due to the rotation of the Earth, the ocean level at the equator is 21 km higher than at the pole. The effect is similar to the curvature of the water surface in a rotating vessel. Similarly, centripetal acceleration acts on the atmosphere. If the atmosphere did not rotate with the Earth, the atmospheric pressure at the equator would be 20 times less than at the pole. The centripetal force $\omega^2 R$ creates a non-hydrostatic pressure gradient that shifts the air to the equator and forms an equipotential surface with constant ground pressure (Holton 2004).

The equatorial zone acts as walls in a rotating vessel both in the ocean and in the atmosphere, forming an equatorial barrier (EB). With an increase in the mass of air or water in some hemisphere, the excess overflows through the equatorial barrier into the other hemisphere due to the centripetal force. Consequently, even minor fluctuations in the angular frequency can have a noticeable effect on the atmosphere (Sidorenkov, 2004).

It's important to note that the centripetal force includes a vertical component perpendicular to the Earth's surface, which slightly reduces the acceleration due to gravity. This component is taken into account when calculating geopotential height. The overall behavior of the centripetal force is not analyzed separately. For an isothermal atmosphere, the magnitude of the centripetal force, denoted as Fc (h, Tφ) at a height h above the surface at latitude φ, can be expressed as follows:

$$F_c(h, T_\varphi) = \frac{P_0}{R_c T_\varphi} exp\left(-\frac{gh}{R_c T_\varphi}\right) \omega^2 R$$

Where $P_o$ is the value of the surface pressure (Pa), $R_c$ is the specific gas constant of air, $T_\varphi$ is the temperature at latitude φ (°K), g is the acceleration of gravity, ω is the angular frequency of rotation of the Earth, R is the radius of the Earth.

If the temperature $T_\varphi$ is constant in the atmosphere, then $F_c(h,T_\varphi)$ is also constant both along the surface and at any height h as follows from the relation (1). Similarly to the walls of a rotating vessel, the equatorial barrier acts on the atmosphere with a force equal to the centripetal force, so the atmosphere is in equilibrium. The observed surface temperature decreases in the direction from the equator to the pole, respectively, the value of $F_c(h,T_\varphi)$ at the surface increases, as the air density increases. On the contrary, in the upper part of the atmosphere, the value $F_c(h,T_\varphi)$ decreases towards the pole, because the exponential term in the ratio (1) decreases faster in cold air than in warm air. At some altitude H, the difference between the centripetal force at the equator and the centripetal force at latitude φ changes sign. For an isothermal atmosphere, according to the ratio (1) H is equal to

$$H = \frac{R_c T_\varphi}{g} \frac{T_0}{T_0 - T_\varphi} ln \frac{T_0}{T_\varphi} \sim \frac{R_c T_\varphi}{g} \qquad (2)$$

Below this height, the local centripetal force is greater than the centripetal force at the equatorial barrier, which causes air to move towards the equator. Above it, the local centripetal force is less than the centripetal force at the equatorial barrier, which causes air to move towards the pole. The value of H in the ratio (1) corresponds to the altitude scale in the atmosphere, which in order of magnitude (8.5 km for the standard atmosphere) corresponds to the height of the polar tropopause (Gavrilova 1982, Ivanova 2013). The change in the direction of the zonal wind occurs at the height of the polar tropopause (Fig. 1), which confirms the correctness of the obtained estimate in relation (2). It is the polar tropopause that is the interface between the surface flow to the equator and the stream flow to the pole.

The centripetal force primarily induces horizontal displacements of air, while vertical motions are necessary for closed circulation patterns. Horizontal flows over the polar tropopause transport air from the equatorial zone to both hemispheres. This leads to a reduction in the total mass of air in the equatorial region and disrupts hydrostatic equilibrium, as the surface pressure should theoretically decrease. Consequently, a vertical pressure gradient is established, resulting in vertical airflow. This airflow is spread across a specific area and supplies air to the horizontal flow. Simultaneously, the horizontal flow over the tropopause transports air to high latitudes, further challenging hydrostatic equilibrium. The surface pressure should theoretically increase. As a result, a downward flow is formed at high latitudes, delivering air to the surface. The regions of descending and ascending air streams are approximately equal in size due to the similar formation mechanisms. Hence, the boundary between them occurs around latitude 30°, dividing the Earth's surface into equal parts. Notably, zones with vertical air movements exhibit a characteristic vertical temperature profile closely resembling the dry-adiabatic profile. This temperature profile can be observed under the tropopause, as shown in Fig.1.

The horizontal flow over the tropopause descends when moving over the interface, which causes its adiabatic heating and forms a tropospheric inversion layer (Birner 2006). Heating forms an inverse temperature profile with respect to the surface. A change in the sign of temperature leads to the appearance of an additional circulation with the opposite movement of air, located above the circulation around the polar tropopause. The processes in the stratosphere have not yet been sufficiently studied to make an estimate of the height of the flow section. It is likely that the height of the flow section can be close to twice the value of the height scale. This corresponds to an altitude of 16-18 km, which corresponds to a change in the direction of the zonal wind at an altitude of 18 km in Fig. 1.

## 4. Structure of flows in the general circulation of the atmosphere

The upper-air sounding data provides valuable insights into the overall atmospheric circulation and suggests the presence of two distinct circulation cells as shown in Fig. 2.

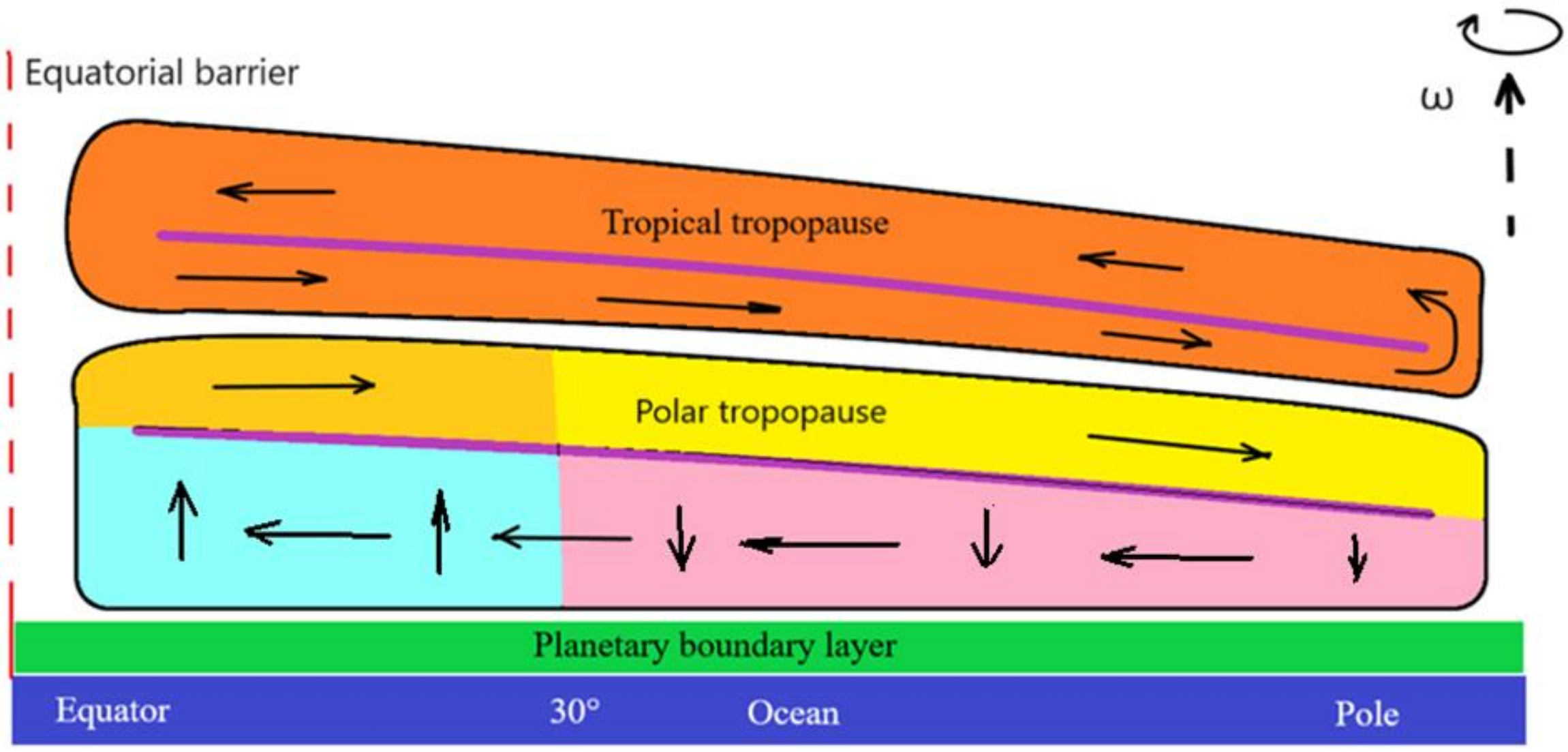

*Fig.2. The structure of flows in the general circulation of the atmosphere. The arrows show the movement of air. The tropopauses are shown as purple lines. The area with a downward flow and air movement from the pole to the equator is highlighted in pink. The region with an updraft and air movement from the pole to the equator is highlighted in blue. The area of flow from the equator to the pole is shown in yellow. The circulation around the tropical tropopause is shown in orange. The planetary boundary layer is shown in green.*

Figure 2 illustrates the flow structure within the global circulation of the atmosphere. The arrows indicate the direction of air movement, while the purple lines represent the location of the tropopauses. The region characterized by a downward flow and air movement from the pole to the equator is highlighted in pink, while the area with an updraft and air movement from the pole to the equator is highlighted in blue. The flow from the equator to the pole is depicted in yellow. The circulation around the tropical tropopause is shown in orange. The planetary boundary layer is depicted in green. The constancy of the tropopause heights is a result of the inertia of the zonal flows. Consequently, local temperature variations, including both daily fluctuations and seasonal changes, are effectively averaged out. Although large-scale temperature changes in the underlying surface can impact the nature of the general circulation, altering the speed of air flows, the fundamental structure remains unchanged. Within the tropopause, strong wind shears and jet streams can be observed due to the transition in wind direction.

Observational data confirms that there are distinct differences in air properties both up to the boundary of the circulatory flow section and above it. In the tropics, moist air rises and undergoes cooling, leading to the condensation of water vapor. Condensation ceases when the relative humidity reaches approximately 60%. Upon entering the stratosphere, the flow maintains a similar humidity level (e.g., as observed in Hawaii, as shown in Fig. 1c). However, the air temperature in the stratosphere is lower by 10-15 degrees compared to the middle latitudes (as depicted in Fig. 1a and Fig. 1b). While the actual water vapor pressure remains constant regardless of temperature, the saturated water vapor pressure increases significantly, which leads to a decrease in relative humidity. An increase in temperature by 10 degrees leads to a fourfold decrease in relative humidity, that is, it decreases to about 15%. With a temperature rise of 15 degrees, the relative humidity increases by seven times, resulting in a relative humidity level of only 8-9%. Consequently, a rapid change in relative humidity is observed at the tropopause.

In the lower atmosphere, up to the level of the polar tropopause, there is a significant concentration of aerosols. However, above the polar tropopause, the presence of aerosols is negligible, resulting in a sudden change in the extinction coefficient of visible light (Kochin 2021). Diffusion processes alone cannot account for the formation of such a distinct boundary with a vertical extent of approximately a hundred meters.

The proposed structure of the GCA aligns with George Hadley's concept of a single-cell circulation between the equator and the poles. The boundary between the ascending and descending air currents is positioned at approximately 30° latitude, consistent with the traditional demarcation between the Hadley and Ferrell cells. In addition, the Ferrell cell at 30° latitude looks like a downdraft, which is also consistent with the proposed model. The presence of an independent circulation cell at temperate latitudes (Ferrell cell) implies a change in the direction of zonal flows, although this is not explicitly evident in the available observational data (as shown in Figs. 1a and 1c). It should be noted that wind data primarily indicate a shift in wind direction between the stratosphere and the troposphere, but provide limited information about the troposphere itself. Detecting the presence of the Ferrell cell within the troposphere alone, if it is an independent cell only in this region, becomes challenging against the backdrop of wind changes attributed to mesoscale processes. The second boundary of the Ferrell cell is situated around 60° latitude, approximately coinciding with the latitude of the Arctic Circle (67°). However, the proposed model focuses solely on the scenario of a temperature minimum at the pole, which does not entirely correspond to the actual processes given the Earth's axial tilt of 23°.

## 5. Conclusion and Implications

The GCA model presented in this paper so far represents only a qualitative description of the phenomena. However, its formulation based on the temperature gradient of the Earth's surface and the centripetal acceleration resulting from the Earth's rotation appears highly convincing due to its alignment with observational findings. Upper-air sounding data provides confirmation that the polar and tropical tropopauses act as boundaries between meridional flows moving in opposite directions. The consistent heights of the tropopauses can be attributed to the flow's inertia, which is influenced by the Coriolis force between these boundaries. A feature of the proposed GCA model is the formation of a vertical temperature profile with a decrease in temperature from the surface to the polar tropopause and a constant temperature above it.

Wind direction change (Mariaccia 2023) is also observed at the stratopause level. This may be due to the existence of another circulation cell in the atmosphere around the stratopause.

GCA plays a crucial role in the transfer of heat and moisture over the Earth's surface with a period of several months, which determines the long-term changes in weather phenomena. To develop a quantitative GSA model, it is necessary to obtain equations for describing the behavior of air in a rotating medium, taking into account the influence of the vertical temperature profile in the atmosphere and the influence of the Coriolis force on the flow velocity.

**Acknowledgments**

The author thanks the staff of the Central Aerological Observatory for their help and useful discussions during the work.